\documentclass[11pt,a4paper]{article}
\pdfoutput=1
\usepackage{jcappub}
\usepackage{graphicx}
\usepackage{amsmath}
\usepackage{physics}
\usepackage{hyperref}
\usepackage{orcidlink}
\usepackage{subcaption}

\newcommand{\Lagr}{\mathcal{L}}

\title{Gravitational waves from a curvature-induced phase transition of a Higgs-portal dark matter sector}

\author[1,2]{Andreas Mantziris \orcidlink{0000-0001-5899-670X},}
\author[1,2]{Orfeu Bertolami \orcidlink{0000-0002-7672-0560}}

\affiliation[1]{Departamento de F\'isica e Astronomia, Faculdade de Ci\^encias, Universidade do Porto, Rua do Campo Alegre 687, 4169-007 Porto, Portugal}

\affiliation[2]{Centro de F\'isica das Universidades do Minho e do Porto, Rua do Campo Alegre s/n, 4169-007, Porto, Portugal.}

\emailAdd{andreas.mantziris@fc.up.pt}
\emailAdd{orfeu.bertolami@fc.up.pt}

\abstract{The study of interactions between dark matter and the Higgs field opens an exciting connection between cosmology and particle physics, since such scenarios can impact the features of dark matter as well as interfering with the spontaneous breaking of the electroweak symmetry. Furthermore, such Higgs-portal models of dark matter should be suitably harmonised with the various epochs of the universe and the phenomenological constraints imposed by collider experiments. At the same time, the prospect of a stochastic gravitational wave background offers a promising new window into the primordial universe, which can complement the insights gained from accelerators. In this study, we examined whether gravitational waves can be generated from a curvature-induced phase transition of a non-minimally coupled dark scalar field with a portal coupling to the Higgs field. The main requirement is that the phase transition is of first order, which can be achieved through the introduction of a cubic term on the scalar potential and the sign change of the curvature scalar. This mechanism was investigated in the context of a dynamical spacetime during the transition from inflation to kination, while also considering the possibility for inducing electroweak symmetry breaking in this manner for a sufficiently low reheating temperature when the Higgs-portal coupling is extremely weak. We considered a large range of inflationary scales and both cases of positive and negative values for the non-minimal coupling, while taking into account the bound imposed by Big Bang Nucleosythesis. The resulting gravitational wave amplitudes are boosted by kination and thus constrain the parameter space of the couplings significantly. Even though the spectra lie at high frequencies for the standard high inflationary scales, there are combinations of parameter space where they could be probed with future experiments.}

\begin{document}

\maketitle

\section{Introduction}

The study of interactions between Dark Matter (DM) and the Higgs field opens interesting prospects for possible investigations and phenomenology of DM through collider experiments and cosmological considerations. The first models \cite{Silveira:1985rk, Bento:2000ah, Burgess:2000yq, Bento:2001yk, Seto:2001ju} along these lines considered an extension of the Standard Model (SM) with an additional singlet scalar field and a portal coupling to the Higgs. Several related models have been studied in the literature \cite{Patt:2006fw, March-Russell:2008lng, Biswas:2011td, Pospelov:2011yp, Mahajan:2012nc, Cline:2013gha, Kouvaris:2014uoa, Costa:2014qga, Duerr:2015aka, Enqvist:2014zqa, Tenkanen:2015nxa, Han:2015hda, Han:2015dua, Bernal:2018kcw, Cosme:2018zbc, Lebedev:2021xey, Arcadi:2024wwg, Cirelli:2024ssz}, with these scenarios being known as ``Higgs-portal" DM. In fact, a putative coupling to dark energy has also been suggested in Refs. \cite{Bertolami:2007wb, Bertolami:2012xn}. Naturally, these set-ups can have an impact on the features of DM itself \cite{Bertolami:2016ywc,Cosme:2017cxk,Cosme:2018nly}, as well as interfering with spontaneous breaking of the electroweak (EW) symmetry \cite{Cosme:2018wfh}. In any case, it is crucial that a model of Higgs-portal DM can be suitably harmonised with the various epochs of our cosmological history and complying with observational and experimental constraints that it inherits.

At the same time, primordial Gravitational Waves (GW) are a ubiquitous feature of the early universe, with various production mechanisms during inflation \cite{Bettoni:2018pbl, Barir:2022kzo, Das:2023nmm, Odintsov:2023weg}, (p)reheating \cite{Kuroyanagi:2017kfx, Artymowski:2017pua, Krajewski:2022ezo, Cosme:2022gwt, Cosme:2022htl, Barman:2023ktz, Barman:2023ymn}, and the radiation-domination epoch \cite{Ellis:2018mja, Ellis:2019oqb, Lewicki:2019gmv, Lewicki:2020azd, Ellis:2020awk, Lewicki:2022pdb, Koutroulis:2023wit}, which cover a large section of the frequency band, since their scales range from Planckian values down to the EW scale \cite{Allahverdi:2020bys, Jinno:2017ixd, Liddle:2000cg}. Since GWs are ripples in spacetime that travel unhindered after emission, any detected GW signature of primordial origin would provide new direct insights into the early universe and the responsible mechanism for its generation. In this sense, cosmological phase transitions have been of increasing interest due to their ability to generate a stochastic GW background that can potentially be detectable with upcoming surveys \cite{LISACosmologyWorkingGroup:2022jok}. In particular, theories Beyond the Standard Model (BSM) offer a promising playground for realising strong First-Order Phase Transitions (FOPT) with prominent GW spectra, where the nature of the ensued phase transition depends on a conjugation of conditions for the background and the parameters of the model under study. While the majority of these considerations revolve around thermal PTs of BSM models at relatively low energy scales of the early universe, with the temperature decrease of the universe after radiation-domination inducing the transition, non-thermal PTs \cite{Cutting:2018tjt, Bettoni:2019dcw, Bettoni:2021zhq, Laverda:2023uqv, Buckley:2024nen} can offer a promising window in the earlier moments of the primordial universe that can complement the insights from different mechanisms and observables.

The purpose of this study was to combine the models developed in Refs. \cite{Cosme:2018wfh} and \cite{Kierkla:2023uzo} to examine whether GWs can be generated from a curvature-induced phase transition of a non-minimally coupled scalar DM model with a portal to the Higgs field. This was explored in a dynamical spacetime during the latest phase of inflation before reheating, while considering the possibility of a curvature-induced breaking of the EW symmetry. The cosmological setting considered was the transition from the inflationary to the kination \cite{Joyce:1996cp} or deflation \cite{Spokoiny:1993kt} era, where the kinetic energy dominates over the potential of the inflaton field. Although an amount of kination is unavoidable in most inflationary models, here we adopt the framework of quintessential-like inflation that is characterised by an extended period of kination, since the inflaton rolls unopposed down the approximately flat tail of its potential after the end of inflation \cite{Peebles:1998qn,Dimopoulos:2001ix,Dimopoulos:2017zvq}. During long kination periods, the total energy density is diluted with the scale factor as $\rho \sim a^{-6}$, resulting in the dominance of any other form of matter over the background \cite{Bettoni:2021qfs}. This behaviour enables an efficient heating of the universe into the standard Hot Big Bang (HBB) epoch through various mechanisms, although not along the usual prescriptions used in the typical oscillatory models and without any direct inflaton-SM couplings,to avoid the destabilisation of the potential's late-time plateau \cite{Bettoni:2021qfs}. The reason for focussing on prolonged kination scenarios, which is also motivated in string cosmology \cite{Conlon:2024uob}, is that it amplifies any corresponding stochastic GW background produced at these scales, making it a promising context for studying cosmological phase transitions \cite{Kierkla:2023uzo}.

The outline of the paper is as follows. In Sec. \ref{sec:2-1}, we present an overview of the cosmological setting for our model, following the evolution of the non-minimally coupled spectator Higgs-portal scalar field from inflation to kination. The dynamical features that induce the phase transition and EW symmetry breaking for both cases of negative and positive non-minimal coupling are discussed in Sec. \ref{sec:2-2}, and Sec. \ref{sec:2-3} includes the necessary consistency analysis for the parameter space of the couplings. The theoretical framework for computing GW signals from bubble collisions due to a FOPT in vacuum is presented in Sec. \ref{sec:3}, with the resulting spectra of the parameter space specified in Sect. \ref{sec:3-3}. In Section \ref{sec:4}, we present our conclusions for the GW signatures obtained, together with the phenomenological constraints imposed that link our model to DM and the breaking of the EW symmetry. In Appendix \ref{app:1}, we offer an overview of the self-interacting massless Higgs-portal DM model of Ref. \cite{Cosme:2018wfh}, as a reference point for comparison. Finally, Appendix \ref{app:2} addresses the necessary transformation of the scalar potential for $\xi_{\phi}<0$, so that we can apply the standard formalism of cosmological phase transitions of Sec. \ref{sec:3}, when the true vacuum of the PT lies at the origin.

\section{Cosmological evolution of a spectator scalar field in the early Universe} \label{sec:2}

\subsection{Non-minimally coupled scalar in a quintessential-like inflationary context}  \label{sec:2-1}

Following the motivation of Ref. \cite{Cosme:2018wfh}, we investigate the cosmological implications of a dark sector with a portal to the Standard Model via the Higgs field. In particular, we are interested in the breaking of the electroweak symmetry via a non-thermal mechanism but also exploring whether there can be any gravitational wave signatures associated with such models. Therefore, we extend the SM with the addition of a massless self-interacting scalar field, $\phi$, which as discussed in Ref. \cite{Cosme:2018wfh}, plays the role of dark matter once it acquires a mass after EW symmetry breaking (EWSB) and oscillates around the minimum of its potential (see Sec. \ref{sec:2-2}), and a decoupled inflaton field $\varphi$ that drives the accelerated expansion of the Universe. This theory consists of two scalar fields, the SM Higgs, $h$, and the dark scalar, which are coupled to each other via the Higgs portal coupling, $g$, and non-minimally coupled to gravity. The latter terms are radiatively generated and required for the renormalizability of the stress-energy tensor in curved spacetime \cite{Birrell:1982ix}, and we cannot disregard the non-minimal couplings $\xi_{h}, \, \xi_{\phi}$ by setting them to zero, as they will become finite at some other renormalization scale due to their running \cite{Mantziris:2020rzh}. It is intuitive to add a renormalizable cubic term for the BSM scalar, which can easily shape the potential into a double-well that could lead to strong FOPTs in many contexts (see e.g. Ref. \cite{Paramos:2002pe}). The full action is given by
\begin{align}
    S=\int d^4 x \sqrt{-g} \left[\frac{M_P^2 - \xi_{\phi} \phi^2 -\xi_h h^2}{2} R-\frac{1}{2}\partial_\mu \phi \partial^\mu \phi- \frac{g^2}{4} h^2 \phi^2 + \frac{\sigma}{3}\phi^3-\frac{\lambda_{\phi}}{4}\phi^4 - \Lagr_h - \Lagr_{\varphi} \right] \,,
    \label{eq:scalar-action}
\end{align}
where $M_P = 2.435 \times 10^{18}$ GeV is the reduced Planck mass, $\sigma$ is the dimensionful cubic coupling, $\lambda_{\phi}$ is the dark scalar self-interaction, $\Lagr_h$ is the Higgs Lagrangian, and $\Lagr_{\varphi}$ is the inflaton Lagrangian density. All couplings are positive, except for $\xi_{\phi}$ that can have either sign, but note that the cubic term with the opposite sign from the others places the additional vacuum of the double-well potential at positive field values.

 The interplay of the scalar potential described above with the time-dependent scalar $R$ from the non-minimally coupled term can lead to the production of GW signals from a curvature-induced phase transition, as in Ref. \cite{Kierkla:2023uzo}. In order to keep our setup as simple as possible, we limit our model to renormalizable terms only with no higher-order factors. However, note that even though there is no $\mathcal{Z}_2$ symmetry to prohibit linear terms of $\phi$ in the BSM potential, we avoid including them, since they can be eliminated by a suitable redefinition \cite{Adams:1993zs}. Any dark scalar-inflaton terms would be irrelevant for the purposes of this study since they would not affect the difference between the vacua for the phase transition, only the location of the potential's minima, according to the considerations discussed in Sec. \ref{sec:2-2}. Higgs-inflaton terms can also be disregarded in this context, given that during inflation $h=0$ and by the time of EW symmetry breaking, the inflaton has a negligibly small value. We highlight that we do not specify an inflationary model in $\Lagr_{\varphi}$, to be as model-independent as possible, and we just utilise a generic parametrisation of quintessential-like inflation, as detailed below. Thus, we assume that the decoupled inflationary sector behaves in the usual manner that provides the necessary exponential expansion of spacetime in the early universe, without including any inflaton-curvature terms in Eq. (\ref{eq:scalar-action}) for the sake of brevity. 

The equation of motion of the inflationary sector, $\Lagr_I = \frac{1}{2}\partial_\mu \varphi \partial^\mu \varphi - U_I$, in Friedmann– Lemaître–Robertson–Walker (FLRW) spacetime is given by 
\begin{align}
    \Ddot{\varphi} + 3H \Dot{\varphi} + \frac{d U_{I}}{d \varphi} = 0 \,,
\end{align}
where $H=\frac{\dot{a(t)}}{a}$ is the Hubble rate, $a(t)$ is the scale-factor of the expanding spacetime, and $U_I$ is the inflaton's potential. The corresponding energy density and pressure of the perfect cosmological fluid are given by
\begin{align}
     \rho_{\varphi}&=\frac12 \Dot{\varphi}^2 + U_{I}(\varphi) \, , \\
     p_{\varphi}&=\frac12 \Dot{\varphi}^2 - U_{I}(\varphi) \, ,
\end{align}
respectively, and the dot denotes a derivative with respect to time. These quantities can be parameterised by the Equation of State (EoS) parameter $w(t) \equiv \frac{p_{\varphi}}{\rho_{\varphi}}$ and by solving the Friedmann and Raychaudhuri equations
\begin{align}
    H^2 &= \frac{\rho_{\varphi}}{3 M_p^2} \,, \\
    \Dot{H} + H^2 &= - \frac{1}{6 M_p^2} \left(\rho_{\varphi} +  3 p_{\varphi}\right) \,,
\end{align}
we can express the first slow-roll parameter in terms of the EoS parameter
\begin{align}
    \epsilon_H \equiv - \frac{\Dot{H}}{H^2}= \frac{3}{2} \left[1 + w(t)\right] \, .
    \label{eq:slow-roll-param}
\end{align}
This differential equation can be solved once the dynamics of the EoS are specified.

\begin{figure}[h!]
    \centering
    \includegraphics[scale=0.9]{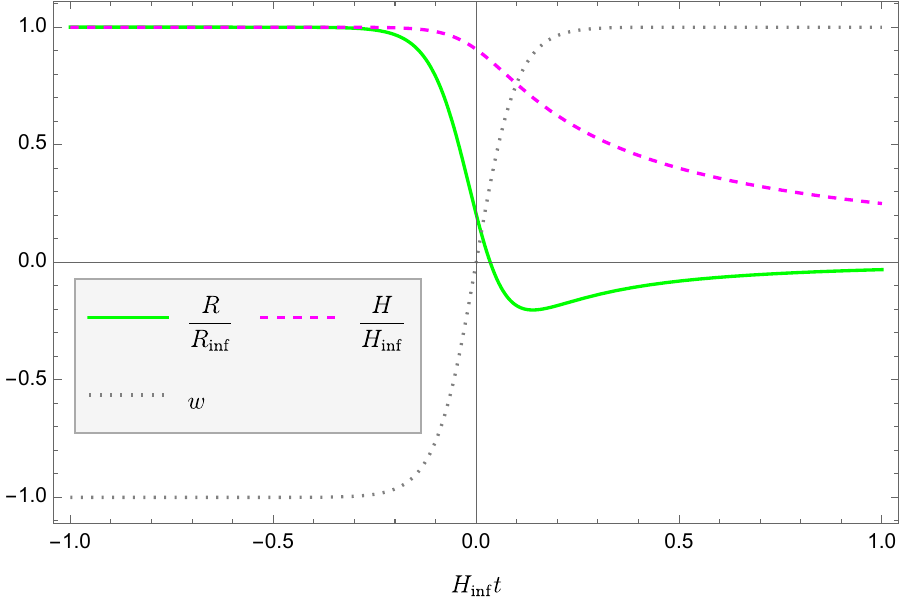}
    \caption{Evolution of the EoS parameter (\ref{eq:w}), the Ricci scalar (\ref{eq:R}), and the Hubble rate (\ref{eq:H/Hinf}) from inflation to kination in Hubble time, with $\beta_w = 10 H_{\rm inf}$.}
    \label{fig:w-R-H}
\end{figure}

As inflation approaches its end, the kinetic term dominates the potential in the energy density of the inflaton, in contrast to the slow-roll regime. This results in an epoch with different dynamics, nested between inflation and the usual radiation-domination period of the HBB era \cite{Bettoni:2021qfs}. When focusing on the transition from inflation to kination, a convenient parametrisation for the barotropic parameter that exhibits a smooth and continuous function linking the de Sitter value $w_{\rm inf}=-1$ with $w_{\rm kin}=1$ during kination \cite{Kierkla:2023uzo}, without being restricted to a particular $U_I$, would be
\begin{align}
    w(t)=\tanh{\left[\beta_w (t-t_0)\right]} \, ,
    \label{eq:w}
\end{align}
where the transition is centred at $t_0$ as shown in Fig. \ref{fig:w-R-H}, the free parameter $\beta_w > 0$ controls its speed and can be adjusted accordingly to match specific inflationary models\footnote{In general, we cannot reverse-engineer to obtain the inflationary model from the EoS parameter without some additional information about the inflaton's behaviour, as $ U_{I}(\varphi, t)=\frac{\dot{\varphi}^2}{2} \frac{w(t)-1}{w(t)+1}$.}. In this treatment, $w$ tends asymptotically to $w_{\rm kin}=1$ mimicking quintessential-like inflationary models, where the inflaton does not oscillate around a minimum, but instead rolls unobstructed along an additional plateau around the scale of the cosmological constant \cite{Bettoni:2021qfs}. In this case, the Hubble rate can be calculated analytically from Eq. (\ref{eq:slow-roll-param}) as
\begin{align}
    \frac{H(t)}{H_{\textrm{inf}}}= \frac{2}{3} \left[H_{\rm inf}(t-t_0) + \left( \frac{H_{\rm inf}}{\beta_w} \right) \ln{\left[ 2 \cosh{\left( \frac{\beta_w}{H_{\rm inf}} H_{\rm inf} (t-t_0) \right)} \right]} + \frac{2}{3} \right]^{-1}\, ,
    \label{eq:H/Hinf}
\end{align}
where $H_{\rm inf}$ is the scale of inflation, and we set $t_0 = 0$, meaning that roughly $t<0$ corresponds to inflation and $t>0$ to kination \cite{Kierkla:2023uzo}. Finally, it is important to consider the cosmological evolution of spacetime curvature in this setting, since the Higgs and the dark scalar are non-minimally coupled to it. A generic expression for the curvature scalar in terms of the EoS parameter is given by
\begin{align}
    R(t)=3\left[1-3w(t) \right]H^2(t) \,,
    \label{eq:R}
\end{align}
where its capability for symmetry breaking mechanisms is manifestly evident by noticing that it switches signs due to evolution of the barotropic parameter $w$, as depicted in Fig. \ref{fig:w-R-H}. This feature of the Ricci scalar makes it a core ingredient of the phase transition triggering mechanism utilised in this study, which was proposed originally in Ref. \cite{Kierkla:2023uzo}. 

\subsection{Curvature-induced phase transitions and electroweak symmetry breaking} \label{sec:2-2}

In this work, we are interested in the cosmological implications of the dark scalar field after inflation, and specifically on its dynamic potential from Eq. (\ref{eq:scalar-action}), which is written as
\begin{align}
    V(\phi, h, R) = \frac{1}{2} \left(\xi_{\phi} R + \frac{g^2}{2} h^2 \right) \phi^2-\frac{\sigma}{3}\phi^3+\frac{\lambda_{\phi}}{4}\phi^4 \,,
    \label{eq:V-potential}
\end{align}
where we have included the Higgs-portal term that it shares with the Higgs potential and the explicit dependence on the curvature scalar. The form of the BSM potential in Eq. (\ref{eq:V-potential}) is chosen as such so that the curvature-induced PT mechanism of Ref. \cite{Kierkla:2023uzo} is used to explore the GW signatures of the oscillating DM model of Ref. \cite{Cosme:2018wfh}, with the minimal addition of a cubic term to the potential so that it possesses a double-well shape that can readily provide a FOPT. To simplify the description of the vacuum transition, we use the semi-analytic expressions from Ref. \cite{Adams:1993zs}, where any fourth-order polynomial potential can be written in the reduced dimensionless form as
\begin{align}
    \widetilde{V}(\Tilde{\phi}, \delta) = \frac{\delta}{2}\Tilde{\phi}^2 - \Tilde{\phi}^3 + \frac{1}{4}\Tilde{\phi}^4 \, .
\end{align}
With this prescription, the dark scalar potential \eqref{eq:V-potential} resembles the form above when the field redefinition and its corresponding quadratic coupling are given by
\begin{align}
\Tilde{\phi} &= \frac{3\lambda_{\phi}}{\sigma}\phi \, , \\
\delta(h, R) &= \frac{9\lambda_{\phi}}{\sigma^2} \left(\xi_{\phi} R + \frac{g^2}{2} h^2 \right) \, ,
\label{eq:delta}
\end{align}
with the overall rescaling of the potential being $V = \frac{(\sigma/3)^4}{\lambda_{\phi}^3} \Tilde{V}$. The cubic term gives a double-well shape to the potential, as long as $ 4 \delta / 9 < 1$,  with three extrema at
\begin{align}
   \phi_1 &= 0 \,, \qquad \qquad \qquad \qquad \qquad \qquad \qquad \qquad \quad \Tilde{\phi}_1 = 0 \, , \label{eq:phi1} \\
   \phi_2 (h, R) &= \frac{\sigma}{2\lambda_{\phi}}\left(1+\sqrt{1 - \frac{4 \lambda_{\phi} \left(\xi_{\phi} R + \frac{g^2}{2} h^2 \right)}{\sigma^2}}\right) \,, \quad \Tilde{\phi}_2 (\delta) = \frac{3}{2} \left( 1 + \sqrt{1 -\frac{4 \delta}{9}} \right) \, , \label{eq:phi2}\\
   \phi_3 (h, R) &= \frac{\sigma}{2\lambda_{\phi}}\left(1-\sqrt{1 - \frac{4 \lambda_{\phi} \left(\xi_{\phi} R + \frac{g^2}{2} h^2 \right)}{\sigma^2}}\right) \,, \quad \Tilde{\phi}_3 (\delta) = \frac{3}{2} \left( 1 - \sqrt{1 -\frac{4 \delta}{9}} \right) \, . \label{eq:phi3}
\end{align}

The evolution of the potential is encoded in the time dependence of the Ricci scalar $R(t)$, and thus the parameter $\delta$ effectively controls the phase transition. In particular, the range $0 \leq \delta \leq 2$ corresponds to configurations between a vanishing barrier and two degenerate minima. However, note that the value of $\delta$, as the redefinition of the couplings in Eq. (\ref{eq:delta}), is not strictly restricted to this range. Putting together Eqs. (\ref{eq:R}) and (\ref{eq:delta}), after disregarding the Higgs portal term\footnote{When the Higgs field acquires a non-vanishing vacuum expectation value (vev), Eq. (\ref{eq:delta_C}) admits an additional term $9 \lambda_{\phi} g^2 v^2 / 2 \sigma^2$. This is negligible for the relevant parameter space and in any case irrelevant for the present calculation, since the FOPTs here take place before EW symmetry breaking.}, allows expressing $\delta$ in a ``normalised'' form,
\begin{align}
    \delta = C \left[ \frac{1 - 3 w(t)}{2} \right] \left( \frac{H(t)}{H_{\rm inf}}\right)^2 \; , \; \; C = 54 \lambda_{\phi} \xi_{\phi} \left( \frac{H_{\rm inf}}{\sigma} \right)^2 \, ,
    \label{eq:delta_C}
\end{align}
where the cosmological evolution is manifestly evident and $C$ is the only free parameter for a given inflationary scale. The potential possess an unbroken state (only one global minimum) when $\delta > 9/4$ and the couplings therefore obey $C \left[ 1 - 3 w(t) \right]  > \frac{9}{2} \left( \frac{H_{\rm inf}}{H(t)}\right)^2$, i.e.
\begin{align}
    C &> \frac{9}{8} \;\;\;\;\;\;\;\;\;\;\;\;\;\;\;\;\;\;\;\;\;\;\;\;\;\;\;\;\;\;\; \mathrm{ during \; inflation \, ,} \\
    C &< - \frac{9}{4} \left( \frac{H_{\rm inf}}{H_{\rm kin}}\right)^2 \approx - \frac{9}{2} \;\;\; \mathrm{ during \; kination \, .}
\end{align}

\begin{figure}[h!]
\begin{minipage}{.5\linewidth}
\centering
\subfloat{\label{fig:Vinf1}\includegraphics[scale=.55]{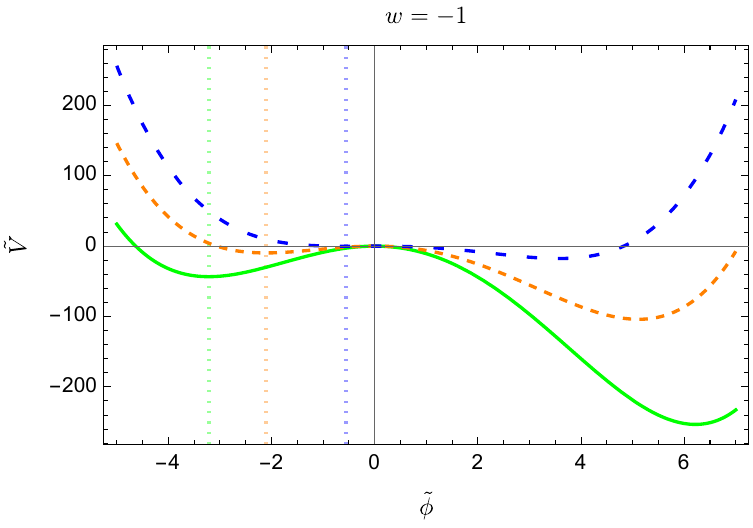}}
\end{minipage}
\begin{minipage}{.5\linewidth}
\centering
\subfloat{\label{fig:Vinf2}\includegraphics[scale=.55]{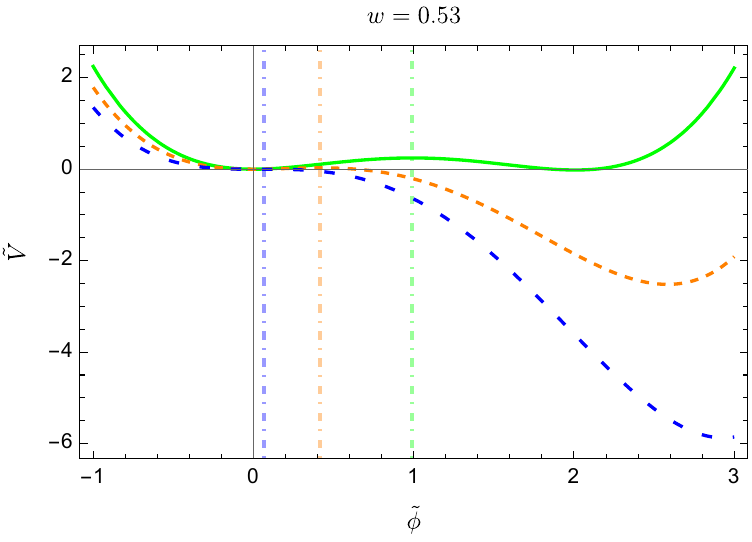}}
\end{minipage} \par\medskip
\begin{minipage}{.5\linewidth}
\centering
\subfloat{\label{fig:Vinf3}\includegraphics[scale=.55]{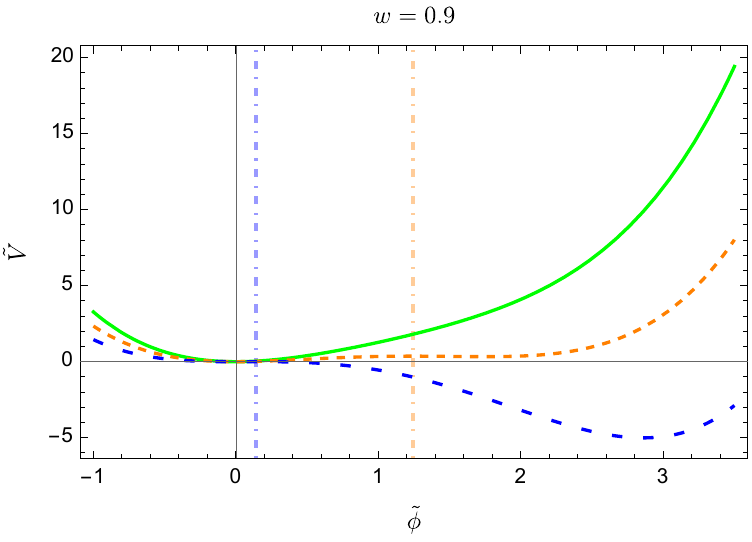}}
\end{minipage}
\begin{minipage}{.5\linewidth}
\centering
\subfloat{\label{fig:Vinf4}\includegraphics[scale=.55]{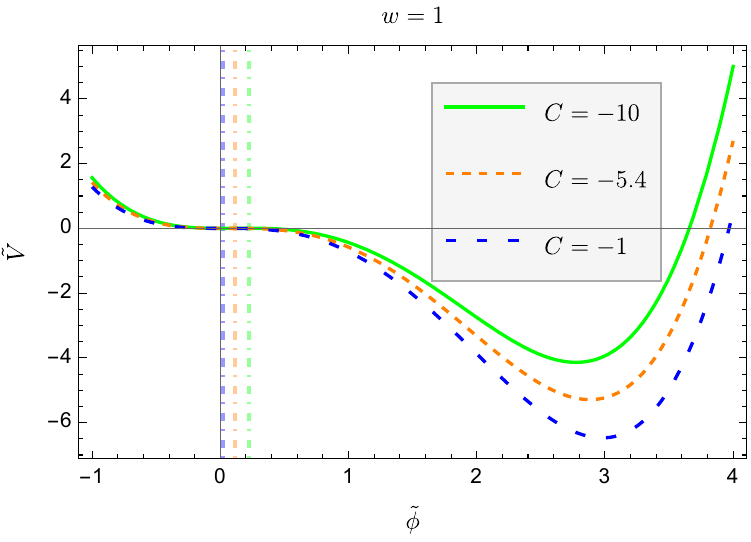}}
\end{minipage}
\caption{Evolution of the reduced dimensionless dark scalar potential for $\xi_{\phi}<0$, from inflation (upper left) to kination (lower right), where the vertical dotted and dash-dotted lines follow the extrema $\Tilde{\phi}_3$ from being the corresponding false vacua to barriers, respectively.}
\label{fig:V-xi-negative}
\end{figure}

Having established a useful framework to express the BSM potential in a convenient parametrisation, we can study its evolution as inflation comes to an end and kination commences. We can identify the minima and maxima of the potential according to the values of its second derivative with respect to the field at the extrema (\ref{eq:phi1})-(\ref{eq:phi3}),
\begin{align}
    \Tilde{V}''(\Tilde{\phi}_1) &\propto \delta \, , \\
    \Tilde{V}''(\Tilde{\phi}_2) &> 0 \; \mathrm{for} \; \delta < 9/4 \, , \\
    \Tilde{V}''(\Tilde{\phi}_3) &\geq 0 \; \mathrm{for} \; \delta \leq 0 \, , \\
    \Tilde{V}''(\Tilde{\phi}_3) &< 0 \; \mathrm{for} \; 0 < \delta < 9/4 \, .
\end{align}
For $\xi_{\phi}<0$, while $R>0$ during inflation, $\delta$ is negative and thus $\Tilde{\phi}_3$ corresponds to the false vacuum (local minimum), $\Tilde{\phi}_2$ to the true vacuum (global minimum), and $\Tilde{\phi}_1$ to the top of the potential barrier (local maximum), as illustrated in Fig. \ref{fig:V-xi-negative}. As $R$ decreases, the increasing $\delta$ reaches zero, where there is an inflection point at $\Tilde{\phi}_3 = \Tilde{\phi}_1 = 0 $ and a global minimum at $\Tilde{\phi}_2 = 3$. During kination, $R<0$ and $\delta > 0$, so that $\Tilde{\phi}_3$ tracks the top of the barrier and $\Tilde{\phi}_1$ is the new false vacuum, while $\Tilde{\phi}_2$ continues to lie at the true vacuum, until the minima become degenerate $\Tilde{V}(\Tilde{\phi}_1)=\Tilde{V}(\Tilde{\phi}_2)=0$ at $\delta=2$, as in the upper right panel of Fig. \ref{fig:V-xi-negative} for $C=-10$. When reaching the threshold $\delta = 9/4$, the potential has an inflection point at $\Tilde{\phi}_2=\frac{3}{2}$ and for $\delta>9/4$ the potential has just a minimum at $\Tilde{\phi}_1=0$, as shown in the lower left panel of Fig. \ref{fig:V-xi-negative} for the $C=-10$ case. Since the curvature scalar will oscillate or tend asymptotically to zero for later times, depending on the inflationary model, the potential will resemble the shape for $\delta=0$, which is shown in the lower right plot of Fig. \ref{fig:V-xi-negative}.

According to the above description, the behaviour of the dark scalar is as follows. During inflation, it has a large vacuum expectation value at $\phi_{\rm inf} = \phi_2$, which contributes to the effective Higgs mass via the Higgs portal, 
\begin{align}
    m_h^{\rm eff} = \sqrt{\xi_h R + \frac{g^2}{2}\phi_{\rm inf}^2} = \sqrt{ \xi_h R + \frac{g^2 \sigma^2}{8 \lambda_{\phi}^2}\left(1+\sqrt{1 - \frac{4 \lambda_{\phi} \xi_{\phi} R}{\sigma^2}}\right)^2} \, ,
\end{align}
and can enhance the stability of the Higgs vacuum at the origin, in addition to the non-minimal coupling of the Higgs, by suppressing its de Sitter fluctuations, 
\begin{align}
   \frac{\langle h^2 \rangle}{\left( H_{\rm inf} / 2 \pi \right)^2} \simeq \frac{H_{\rm inf}}{m_h^{\rm eff}} = \frac{ H_{\rm inf} / \sigma}{\sqrt{12 \xi_h \left(\frac{H_{\rm inf}}{\sigma} \right)^2 + \frac{1}{2} \left(\frac{g}{\lambda_{\phi}}\right)^2 \left(\frac{1}{2} + \frac{1}{2} \sqrt{1 - 48 \lambda_{\phi} \xi_{\phi}  \left(\frac{H_{\rm inf}}{\sigma} \right)^2}\right)^2}} \, ,
\end{align}
if $g>\lambda_{\phi}$ within the relevant parameter space of the couplings $\sigma \sim H_{\rm inf} \, , \; \lambda_{\phi} < 10^{-3} \, , \; \xi_h \gtrsim 0.1$\footnote{In fact, the Higgs curvature coupling can be much larger, $\xi_{h} \sim \mathcal{O} \left( 1 - 10^2\right)$, if we account for the survival of the EW vacuum after inflation \cite{Laverda:2024qjt, Li:2022ugn, Figueroa:2021iwm, Figueroa:2017slm}, and thus it can solely suppress the fluctuations decisively.} \cite{Cosme:2018wfh, Kierkla:2023uzo, Mantziris:2022fuu}. During kination, for a subset of the parameter space of the BSM couplings, it is possible for the dark scalar to tunnel through the potential barrier from $\phi_2$ to $\phi_1=0$, as shown in the lower left panel of Fig. \ref{fig:V-xi-negative} for $C=-5.4$. If this process happens quickly enough with respect to the evolution of the background, it can result in a FOPT with an accompanying production of $\phi_1$-bubbles expanding in a $\phi_2$-universe. The collisions of these bubbles in the cold vacuum of the pre-HBB universe produce GW spectra, which are characterised by the transition strength and duration, as described in Sec. \ref{sec:3}.

In parallel to the BSM field, we have to take into account the dynamics of the spectator Higgs field, since it is coupled both to the dark scalar via the Higgs-portal and to the curvature scalar. Following the evolution of the effective Higgs potential in this cosmological setting is a complicated and non-trivial matter, due to the renormalization scale dependence of the SM parameters, pole-matching, loop and curvature corrections, renormalization group improvement techniques, and higher-order operators \cite{Glavan:2023lvw, Steingasser:2023ugv, Mantziris:2022fuu,Cruz:2022ext, Mantziris:2020rzh,Ema:2020evi, Ai:2020sru, Hardwick:2019uex, Fumagalli:2019ohr, Markkanen:2018pdo, Markkanen:2018bfx, Rajantie:2017ajw, Markkanen:2017dlc,Rajantie:2016hkj, Herranen:2014cua, Chetyrkin:2012rz,Ford:1992mv}. This is beyond the scope of this study, and therefore we derive our conclusions considering the simplest form of the Higgs potential at tree-level, $U_H = \frac{-m^2_h + \xi_h R}{2} h^2 + \frac{\lambda_h}{4} h^4$. This serves as an adequate approximation, given that the phase transition of the BSM field proceeds much faster than the evolution of the Higgs potential and the background's. Hence, for the central values of the SM parameters $\lambda_h < 0$ \cite{ParticleDataGroup:2022pth} and the true vacuum of the Higgs field during inflation lies at
\begin{align}
    h = \sqrt{v^2 + \frac{\xi_h R}{|\lambda_h|} + \frac{g^2 \phi^2}{2 |\lambda_h|}} \, ,
    \label{eq:h_EW}
\end{align}
where $v = \frac{m_h}{\sqrt{\lambda_h}} \simeq 246.22$ GeV is the Higgs' vev at the EW scale. After inflation, the BSM phase transition takes place and the dark scalar is found at $\phi = \phi_1=0$. We assume that the reheating temperature is sufficiently low, $T_{\rm reh}  \leq 80$ GeV, so that the EW symmetry breaking cannot be induced thermally, as proposed in Ref. \cite{Cosme:2018wfh}. However in our case, EWSB is induced by the decrease of the curvature scalar after crossing the threshold
\begin{align}
    R_{\rm th} = \frac{|\lambda_h|}{\xi_h} v^2 \, ,
    \label{eq:R_th}
\end{align}
unlike Ref. \cite{Cosme:2018wfh}, where it is induced post-inflation by the dark scalar rolling down its potential, as described in Appendix \ref{app:1}. Shortly after the PT, the dark scalar potential develops a high-field vev $\phi_2 \approx \frac{\sigma}{2 \lambda_{\phi}}$, without a potential barrier to separate the two vacua. Therefore, since there are no stabilising or symmetry restoration mechanisms, it is inevitable that the dark scalar will roll towards the true vacuum in a manner resembling a second-order phase transition. In this case, we do not expect strong GW signatures to be generated, but the EW symmetry could be restored, unless the portal coupling is sufficiently small\footnote{For phenomenological flexibility, we quote the weakest bound on the portal coupling, corresponding to the largest Higgs self-coupling $\lambda_h (\mu_{\rm EW}) \approx 0.13$ \cite{Mantziris:2020rzh}, but our calculation is technically performed at $\mu_{\rm inf} \gg \mu_{\rm EW}$.},
\begin{align}
    g < \sqrt{\frac{2 |\lambda_h| \lambda_{\phi}}{\sigma}} v < 125 \sqrt{\frac{\lambda_{\phi}}{\sigma}}\, ,
    \label{eq:g-bound}
\end{align}
so that it suppresses the large vev of the BSM scalar, $\phi_{\rm kin}=\phi_2 \propto \frac{\sigma}{\lambda_{\phi}}$.

\begin{figure}[h!]
\begin{minipage}{.5\linewidth}
\centering
\subfloat{\includegraphics[scale=.55]{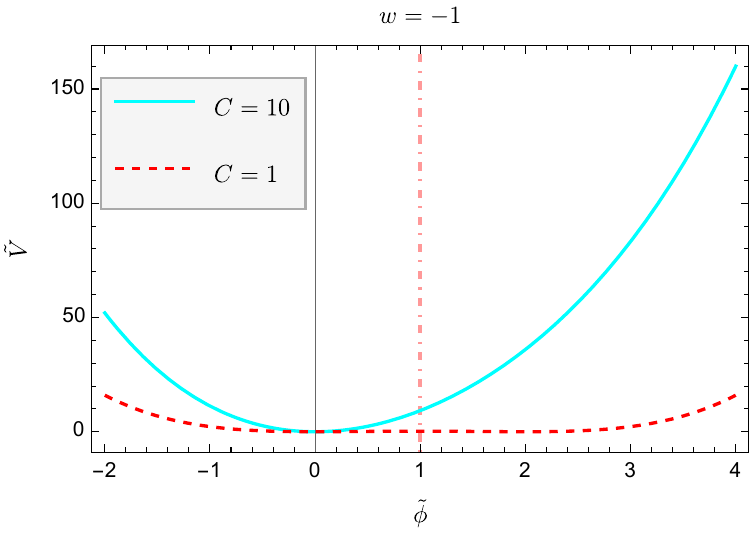}}
\end{minipage}
\begin{minipage}{.5\linewidth}
\centering
\subfloat{\includegraphics[scale=.55]{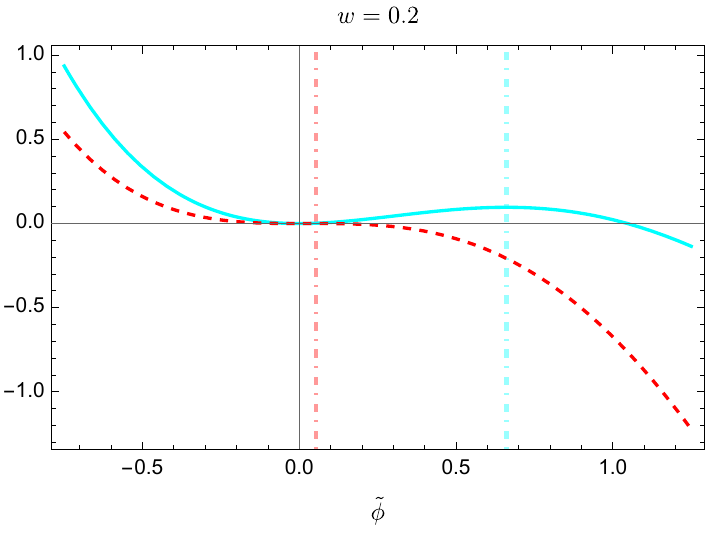}}
\end{minipage} \par\medskip
\begin{minipage}{.5\linewidth}
\centering
\subfloat{\includegraphics[scale=.55]{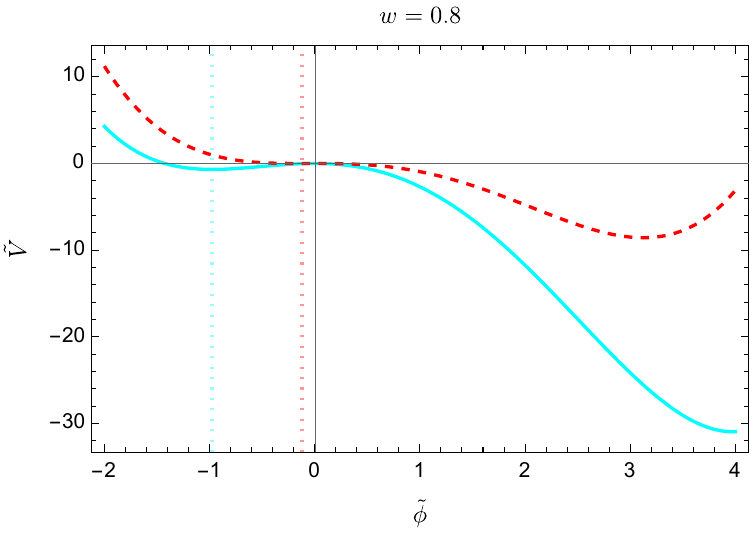}}
\end{minipage}
\begin{minipage}{.5\linewidth}
\centering
\subfloat{\includegraphics[scale=.55]{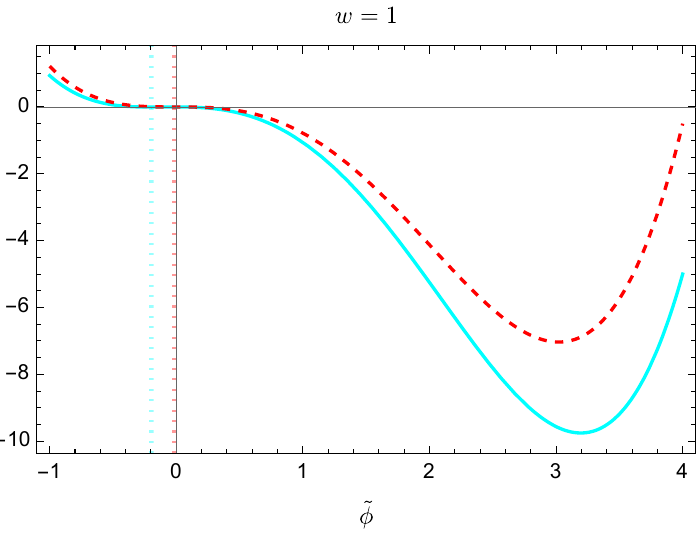}}
\end{minipage}
\caption{Evolution of the BSM potential for $\xi_{\phi}>0$, in the reduced dimensionless form, from inflation (upper left) to kination (lower right). The vertical dash-dotted and dotted lines follow the extrema $\Tilde{\phi}_3$ from being the corresponding barriers to false vacua, respectively.}
\label{fig:V-xi-positive}
\end{figure}

The situation is different for positive non-minimal coupling $\xi_{\phi}>0$. During inflation, the dark scalar lies in its true vacuum $\Tilde{\phi}_1$ at the origin, with the potential being symmetric or having a false vacuum state at higher field values, depending on the couplings \cite{Kierkla:2023uzo}. As inflation ends and kination proceeds, a deeper vacuum state $\Tilde{\phi}_2$ develops with a decreasing potential barrier between the two located at $\Tilde{\phi}_3$, as shown in Fig. \ref{fig:V-xi-positive}. When the barrier becomes thin enough, the field can tunnel through resulting in a FOPT and the nucleation of $\phi_2$-bubbles. If the PT proceeds fast enough, the bubbles percolate and GWs are produced. Since the dark scalar has acquired a large vev, the EW symmetry cannot be broken unless the portal is weaker that the bound of Eq. (\ref{eq:g-bound}). Therefore, adopting the same assumption for a low reheating temperature $T_{\rm reh} \leq 80$ GeV \cite{Cosme:2018wfh}, EWSB is again induced by the evolving curvature crossing the threshold (\ref{eq:R_th}).

\subsection{Consistency analysis and phenomenological considerations} \label{sec:2-3}

Finally, we close this section with a discussion on the constraints that the BSM scalar has to obey to comply with the assumptions of our cosmological scenario and particle interactions. Firstly, the dark scalar must behave as a spectator field at early times without affecting the inflationary dynamics, $|V(\phi_{\rm inf}, 0, R = 12 H_{\rm inf}^2)| < 3 M_P^2 H^2_{\rm inf}$. For $\xi_{\phi}<0$, $\phi_{\rm inf} = \phi_2$ and thus
\begin{align}
     \frac{\sigma^2}{\lambda^3_{\phi}} \left(1- \frac{4 C}{3} + \sqrt{1 - \frac{8 C}{9}} \right) \left( 1 + \sqrt{1 - \frac{8 C}{9}}\right)^{2} < 288 M_P^2 \left(\frac{H_{\rm inf}}{\sigma} \right)^2 \, .
     \label{eq:inf_bound}
\end{align}
Three bounds can be inferred that enable the identification of the relevant parameter space of the couplings, namely
\begin{align}
     \lambda^{3}_{\phi} > \frac{1}{36} \left(\frac{\sigma}{M_P}\right)^2\left(\frac{\sigma}{H_{\rm inf}} \right)^2 \quad &\mathrm{for} \quad - 8 C / 9 \ll 1 \, ,\\
      \lambda^{3}_{\phi} > \frac{1}{26} \left(\frac{\sigma}{M_P}\right)^2\left(\frac{\sigma}{H_{\rm inf}} \right)^2 \quad &\mathrm{for} \quad - 8 C / 9 \approx 1 \, ,\\
     {\lambda_{\phi}} > 12  \left(\frac{H_{\rm inf}}{M_P}\right)^2 \xi_{\phi}^2  \quad &\mathrm{for} \quad - 8 C / 9 \gg 1 \, .
\end{align}
In the opposite case of $\xi_{\phi}>0$, $\phi_{\rm inf} = \phi_1 = 0$ and so the dark scalar is by default spectating during inflation, which implies that we cannot infer any constraints on its couplings. 

After EWSB, we assume that for the Higgs branching ratio into invisible particles, the latter correspond to the dark scalar field. Hence, an upper bound can be placed on the portal coupling and the corresponding effective mass \cite{Cosme:2018wfh},
\begin{align}
    g < 0.13 \, , \qquad m_{\phi} \lesssim 22.6 \textrm{ GeV} \, ,
    \label{eq:g_branching_ratio_bound}
\end{align}
where the mass is generated solely by the Higgs portal
\begin{align}
    m_{\phi}=\frac{g}{\sqrt{2}} v \, .
\end{align}
In addition, radiative corrections from the portal to the BSM self-coupling should be suppressed to avoid fine tuning \cite{Cosme:2018wfh},
\begin{align}
    \Delta \lambda_{\phi} \sim \frac{g^4}{16 \pi^2} < \lambda_{\phi} \, .
    \label{eq:g_radiative}
\end{align}
By requiring the portal coupling to satisfy the bound (\ref{eq:g-bound}), Eq. (\ref{eq:g_radiative}) provides an intuitive upper constraint on the BSM couplings that allow for curvature-induced EWSB,
\begin{align}
    \lambda_{\phi} < \left( \frac{4 \pi^2}{\lambda^2_h v^4} \right) \sigma^2 \approx \left( \frac{H_{\rm inf}}{10^{3}} \right)^2 \left(\frac{\sigma}{H_{\rm inf}}\right)^2 \, ,
\end{align}
which is most constrictive for the largest Higgs quartic coupling \cite{Mantziris:2020rzh}, $\lambda_h (\mu_{\rm EW}) \approx 0.13$ . 

Finally, the BSM scalar plays the role of dark matter once it acquires a mass after EWSB, oscillating around its potential minimum, as discussed in Ref. \cite{Cosme:2018wfh}. Thus, the abundance of dark matter $\Omega_{\phi, 0} \simeq 0.26$ dictates the value of the portal coupling with respect to the cubic and quartic BSM couplings,
\begin{align}
    g = \frac{\sqrt{12 \Omega_{\phi, 0}} H_0 M_P}{v} \left( \frac{g_{* \mathrm{reh}}}{g_{*0}}\right)^{\frac{1}{2}} \left( \frac{T_{\mathrm{reh}}}{T_0}\right)^{\frac{3}{2}} \phi_{\rm reh}^{-1} \simeq 3 \times 10^{-17} \left(\frac{T_{\mathrm{reh}}}{80 \, \mathrm{GeV}}\right)^{\frac{3}{2}}  \frac{\lambda_{\phi}}{\sigma}\, ,
    \label{eq:g_DM}
\end{align}
where the Hubble constant is $H_0 \simeq 1.5 \times 10^{-42}$ GeV, the effective degrees of freedom at reheating was $ g_{* \mathrm{reh}} = 106.75$ and at present is $g_{* 0}=3.36$, $\phi_{\rm reh}$ is the amplitude of the BSM scalar at reheating, and the CMB temperature today is $T_0=23.48 \times 10^{-5}$ GeV \cite{Husdal:2016haj, Liddle:2000cg}. In our model, the dark scalar settles into the true vacuum $\phi_{\rm reh} = \phi_2 (R=0,h=v) \approx \frac{\sigma}{\lambda_{\phi}}$ at late times, regardless of the sign of its non-minimal coupling $\xi_{\phi}$. Therefore, the quartic term always dominates the other terms in the potential, and the dark scalar oscillates around $\phi_2$.

\section{Primordial gravitational waves from bubble collisions during kination} \label{sec:3}

\subsection{Framework of cosmological phase transitions in vacuum}

As we saw in the previous section, there is a possibility for the BSM scalar to undergo a FOPT in the early universe because of the evolution of the curvature scalar. As a result, bubbles of true vacuum would nucleate at the points where the field has tunnelled through the potential barrier. These bubbles would grow and quickly reach a relativistic expansion rate due to the energy difference between the vacua inside and outside of the bubble wall \cite{Callan:1977pt, Coleman:1980aw}. In the cold post-inflationary scenario considered here, there is no thermal plasma between the bubbles, and evidently there are no temperature and friction effects, as reheating takes place at much lower scales.

In the context of cosmological phase transitions, the probability of tunnelling from false (fv) to true (tv) vacuum is given by the bubble nucleation rate (per spacetime volume)
\begin{align}
    \Gamma = F \left(\frac{S^E_4}{2\pi}\right)^2 \exp(-S^E_4) \, ,
    \label{eq:Gamma}
\end{align}
where $S_4^E$ is the Euclidean action for the corresponding $O(4)$-symmetric bounce solution of a double-well potential in vacuum \cite{Turner:1992tz, Fairbairn:2019xog, Matteini:2024xvg}. In the same fashion as in Refs. \cite{Kierkla:2023uzo, Ellis:2020awk}, we make use of the semi-analytic approximation for the Euclidean action of the critical bubble,
\begin{align}
    S_4^E (\delta) = \frac{4 \pi^2}{3 \lambda_{\phi}}(2-\delta)^{-3}\left[\alpha_1\delta + \alpha_2 \delta^2 + \alpha_3 \delta^3\right] \, ,
    \label{eq:S4E}
\end{align}
where $\alpha_1 = 14.1304$, $\alpha_2 = -11.1304$, $\alpha_3 = 2.1576$ are numerically fitted parameters \cite{Matteini:2024xvg}, which are updated to the ones derived in Ref. \cite{Adams:1993zs}. This formalism applies to cases where the fv is fixed at the origin and the tv has a negative potential energy with {\color{blue}a positive} vev. In the thin-wall approximation, the prefactor $F$ for cold transitions is given by
\begin{align}
        F_{\rm thin} = R_0^{-4} \approx \frac{\pi^2\Delta V}{2S_4^E} \, ,
        \label{eq:F_thin}
\end{align}
where $\Delta V = V(\phi_{\rm fv}) - V(\phi_{\rm tv})$ is the energy difference between the two vacua, and the bubble-wall radius $R_0$ is estimated after neglecting the energy stored in the bubble wall \cite{Coleman:1977py}. This approach is generally used when the height of the potential barrier is much larger than the energy difference between the two vacua, as for example in Fig. \ref{fig:V-xi-negative}, but in scenarios of vanishing potential barriers as in Fig. \ref{fig:V-xi-positive}, this approximation is no longer valid. In such cases, the standard formalism \cite{Linde:1980tt,Linde:1981zj, Coleman:1977py, Callan:1977pt, Ellis:2018mja} for the estimation of the prefactor in Eq. (\ref{eq:Gamma}) is purposely avoided, because the critical radius is not well defined. Hence, the prefactor is typically estimated with a dimensional approximation involving the field's vevs
\begin{align}
    F_{\rm thick} = \left( \Delta \phi_v \right)^4= \left(\phi_{\rm fv}-\phi_{\rm tv}\right)^4 \, .
    \label{eq:F_thick}
\end{align}

However, a recent study \cite{Matteini:2024xvg} has managed to provide a one-loop result that can adequately interpolate between thin and thick walls, with the corresponding prefactor being 
\begin{align}
    F_{\rm 1-loop} = (m^2)^2 e^{-\frac{1}{2} \Sigma_4} \, ,
    \label{eq:F_both}
\end{align}
where $m^2$ corresponds to the canonically normalised quadratic ``mass'' term in the potential of the scalar field undergoing the phase transition, and the one-loop contribution to the Euclidean action is numerically approximated by the regularised sum over multipoles
\begin{align}
    \Sigma_4 (\delta) = \frac{53.9926 - 47.6801 \delta +11.0134 \delta^2 +0.3358 \delta^3 + 0.4197 \delta^4 - 0.2938 \delta^5}{(2 - \delta)^3} \, .
\end{align}
Hence for our study, we will use Eq. (\ref{eq:F_both}) to derive the resulting GW spectra, since it can be readily applied to both scenarios of positive and negative non-minimal coupling and the numerical computations are more stable. To apply this formalism to the case of $\xi_{\phi}<0$, where the tunnelling takes place from $\phi_2 >0$ to $\phi_1 =0$, we have to perform a redefinition of the potential, $V \rightarrow W$, that fixes the fv at the origin, as shown explicitly in Appendix \ref{app:2}. The corresponding quadratic ``mass'' terms are given according to (\ref{eq:W_potential}) and (\ref{eq:V-potential}) by
\begin{align}
    m^2_{\xi_{\phi}<0} (R, h=0) &= \frac{\sigma^2}{4 \lambda_{\phi}} \left(1 + \sqrt{1 + \frac{4 \lambda_{\phi} |\xi_{\phi}| R}{\sigma^2}}\right) + |\xi_{\phi}| R \, , \\
    m^2_{\xi_{\phi}>0} (R, h=0) &= \xi_{\phi} R \, ,
\end{align}
and we have to use the appropriate $\delta$ in each case, i.e. Eq. (\ref{eq:deltaW}) or Eq. (\ref{eq:delta}), respectively. We included the standard prescriptions of Eqns. (\ref{eq:F_thin}) and (\ref{eq:F_thick}), even though we did not use them for the calculation of the GW spectra in Sec. \ref{sec:3-3}, for the sake of completeness and to provide a reference point for the comparison of these different approaches.

\subsection{Bubble nucleation and phase transition parameters}

Whether bubble production is efficient enough to result in percolation, within a particular cosmological setting, is determined by the nucleation condition
\begin{align}
\int_{t_{\rm col}}^{t_{\rm nuc}} dt \frac{\Gamma(t)}{H(t)^3} = 1 \, , 
\label{eq:nucl}
\end{align}
where $t_{\rm nuc}$ corresponds to the time of bubble nucleation and $t_{\rm col}$ to that of collision \cite{Ellis:2018mja}. In our investigation, any potential FOPTs would be induced during the transition from inflation to kination, where the Hubble parameter changes only slightly, as shown in Fig. \ref{fig:w-R-H}. Therefore, we consider it being approximately constant for the duration of any phase transitions taking place during this period \cite{Kierkla:2023uzo}. In addition, we assume that the PT proceeds and completes much faster than the evolution of the background. This means that the time elapsed from bubble production until collision is sufficiently small compared to cosmic evolution, so that $t_{\rm nuc} \approx t_{\rm col}$\footnote{For all practical purposes in the remainder of this work, we will use these two subscripts interchangeably, even though we are explicitly denoting when each quantity is calculated.} \cite{Ellis:2018mja, Ellis:2019oqb, Kierkla:2022odc}. These assumptions allow us to express the nucleation condition (\ref{eq:nucl}) in a simpler form as
\begin{align}
    \Gamma(t_{\rm nuc})=H_{\rm nuc}^4 \, ,
    \label{eq:nucl2}
\end{align}
which translates to the formation of one true-vacuum bubble per Hubble volume. By inserting the decay rate (\ref{eq:Gamma}) in Eq. (\ref{eq:nucl2}), we can estimate the time of bubble nucleation $t_{\rm nuc}$. However, note that for $\xi_{\phi}<0$, the decay rate is very sensitive to the choice of the BSM couplings, because of the required evolution of the $\phi_2$-vacuum from a true to a short-lived false vacuum. Thus, some amount of fine-tuning is required to extract $t_{\rm nuc}$ effectively.

After the bubbles are formed, they grow rapidly in the post-inflationary universe and fill the universe, with the BSM scalar field being in its true vacuum within a cosmological horizon. During percolation, bubble collisions generate a stochastic GW background, if the PT is strong enough and completes sufficiently quickly \cite{Allahverdi:2020bys, LISACosmologyWorkingGroup:2022jok}. This is effectively quantified by two PT parameters. The first is the inverse duration of the transition $\beta_{\rm col}$, quantifying the inverse of the time interval between bubble formation and percolation, which can be thought of as the size of the bubbles at the time of collision, after assuming exponential bubble expansion \cite{Allahverdi:2020bys, Kierkla:2023uzo}. Depending on the evolution of the Euclidean action, this parameter can be evaluated as 
\begin{align}
    \beta_{\rm col}^{\xi_{\phi}<0} = \eval{\sqrt{ \frac{d^2 S_4^E }{dt^2}}}_{t=t_{\rm col}} \, , 
    \label{eq:beta_star-xi_negative} \\
      \beta_{\rm col}^{\xi_{\phi}>0} = - \eval{ \dv{t} S_4^E }_{t=t_{\rm col}} \, ,
    \label{eq:beta_star-xi_positive}
\end{align}
where in the first case $S_4^E$ reaches its minimum before the transition completes, while in the latter the Euclidean action decreases slowly in time \cite{Cutting:2018tjt}, according to the dynamics of the phase transitions shown in Figs. \ref{fig:V-xi-negative} and \ref{fig:V-xi-positive}. Note that in each case, we have to use the appropriate expression for $\delta$, i.e. Eq. (\ref{eq:deltaW}) for $\xi_{\phi}<0$ and Eq. (\ref{eq:delta}) for $\xi_{\phi}>0$. Since we have assumed that the phase transition is sufficiently quick, so that the background can be considered non-dynamic during the process, the hierarchy between the respective velocities has to be maintained $\beta_{\rm col} > \beta_w = 10 H_{\rm inf}$, which places an implicit constraint on the parameter space of the couplings.

The second parameter is the strength of the PT, which corresponds to the ratio of latent heat released from tunnelling to the lower ground state over the background energy at collision time \cite{LISACosmologyWorkingGroup:2022jok, Ellis:2020awk, Kierkla:2023uzo}, and is given by
\begin{align}
    \alpha \equiv \frac{\rho_{\rm PT}}{\rho_{\rm bac}} \, .
    \label{eq:alpha}
\end{align}
This quantity can be evaluated analytically, as the energy difference between the two vacua \cite{Kierkla:2022odc} is given by
\begin{align}
    |\Delta V|= |V(\phi_2) - V(\phi_1)| = \frac{\sigma^4}{96 \lambda_{\phi}^3} \left( 1 + \sqrt{1- \frac{4 \delta}{9}} \right)^2  \left( 1 - \frac{2 \delta}{3} + \sqrt{1- \frac{4 \delta}{9}} \right) \,,
    \label{eq:DeltaV}
\end{align}
and holds for both cases of positive and negative $\xi_{\phi}$ (invariant under the redefinition of Appendix \ref{app:2}), and given that the inflaton field dominates the energy density and has not decayed during kination, the background energy density is $\rho_{\rm bac} = \rho_{\rm kin} = 3 M_P^2 H^2_{\rm inf}$. Thus, the transition strength reads
\begin{align}
    \alpha = \frac{|\Delta V|}{3 M_P^2 H_{\rm inf}^2} = \frac{1}{72} \left(\frac{H_{\rm inf}}{M_P}\right)^2 \lambda_{\phi}^{-3} \left(\frac{\sigma}{H_{\rm inf}}\right)^4 \left[ 1 + \sqrt{1- \frac{4 \delta}{9}} -\frac{2 \delta}{3} - \frac{4 \delta}{9} \sqrt{1- \frac{4 \delta}{9}}  + \frac{2 \delta^2}{27} \right]  \, ,
    \label{eq:alpha-BSM}
\end{align}
and is evaluated at the time of nucleation. We have purposely written the above expression in increasing orders of $\delta$ to demonstrate its influence on $\alpha$, as we roughly have $\delta_{\xi_{\phi}<0}(t_{\rm nuc}) \approx 1$ and $\delta_{\xi_{\phi}>0}(t_{\rm nuc}) \rightarrow 0$. We restrict ourselves to sufficiently strong FOPTs that can generate amplified GW signals, without overtly large values of $\alpha$ where supercooling and additional considerations must be taken into account \cite{Athron:2022mmm}, i.e. $10^{-4} \lesssim \alpha \lesssim 10^{-1}$.

\subsection{Gravitational wave spectra} \label{sec:3-3}
Any GWs generated in the primordial universe travel unhindered throughout the fabric of spacetime and even tough they were very energetic events at emission, their spectra are getting red-shifted as the universe expands. However, the alternative expansion history of the universe after inflation in scenarios with prolonged kination affects the spectral shape of the GW signal. In particular, the tilt of the spectrum is modified as $\Omega_{\mathrm{GW}}(f) \propto f^4$ for the modes entering the horizon during kination \cite{Gouttenoire:2021jhk}, and reverts back to the usual cubic dependence after reheating \cite{Durrer:2003ja, Caprini:2009fx, Cai:2019cdl}. This asymmetric signature in the GW spectra, as shown also in Ref. \cite{Kierkla:2022odc}, acts as a ``smoking-gun'' feature of an extended period of kination. Therefore, the GW amplitude from a PT induced during kination redshifts until the present day \cite{Allahverdi:2020bys, Kierkla:2023uzo} is given by
\begin{align}
    \Omega_{\rm GW, 0}(f) h^2 = 1.67 \times 10^{-5} \left(\frac{H_{\rm col}}{H_{\rm reh}}\right)^{2\frac{3w_{\rm int}-1}{3w_{\rm int}+3}}  \left(\frac{H_{\rm col}}{\beta_{\rm col}}\right)^{2} \left(\frac{\alpha}{\alpha+1}\right)^2 S(f) \,, 
    \label{eq:GW}
\end{align}
where $w_{\rm int} \approx 1$ corresponds to the integrated value of the EoS parameter over redshift from the time of bubble collision $t_{\rm col}$ to reheating $t_{\rm reh}$, which approximates $w_{\rm kin}=1$ for quintessential inflation-like scenarios. The spectral shape $S(f)$ is defined using a numerically derived broken power law~\cite{Lewicki:2020azd, Lewicki:2019gmv}
\begin{align}
S(f) = 25.10 \, \left[ 2.41 \, \left(\frac{f}{f_{\rm peak, 0}}\right)^{-0.56} + 2.34 \, \left(\frac{f}{f_{\rm peak, 0}}\right)^{0.57} \right]^{-4.2} \,, 
\end{align}
where $f_{\rm col} = 0.13 \beta_{\rm col}$ is the peak frequency of the spectrum at the moment of bubble percolation \cite{Allahverdi:2020bys,Kierkla:2023uzo}, which at present has redshifted to
\begin{align}
     f_{\rm peak, 0} = 1.65 \times 10^{-5} \left(\frac{H_{\rm col}}{H_{\rm reh}}\right)^{\frac{3w_{\rm int}-1}{3w_{\rm int}+3}} \left(\frac{f_{\rm col}}{H_{\rm col}}\right) \left(\frac{T_{\rm reh}}{100 \mbox{ GeV}}\right) \, . 
    \label{eq:fpeak}
\end{align}

We adopt the simplified assumption of instantaneous thermalisation after kination, where the reheating temperature can be related to the corresponding Hubble rate as
\begin{align}
   H_{\rm reh}  = \frac{\pi}{3} \sqrt{\frac{g_{*\mathrm{reh}}}{10}} \frac{T_{\rm reh}^2}{M_{P}} \,.
   \label{eq:H_reh}
\end{align}
This approach purposely avoids dealing with the complicated dynamics of the reheating process, because the primary focus of our work is on the GW spectra, which are only affected directly by the order of magnitude of the reheating scale. Although many different mechanisms with very interesting and rich phenomenology could be explored here, their overall impact on the GW signatures would not be significant. In this work, the reheating temperature is assumed to be $T_{\rm reh}  \leq 80$ GeV \cite{Cosme:2018wfh} so that the EW symmetry is broken by the dynamic curvature, and temperature effects cannot restore it afterwards. Hence, the scale of reheating is estimated at $H_{\rm reh} \approx 9 \times 10^{-15} $ GeV. Finally, we highlight that settings with prolonged kination can typically constrain the extent of reheating, but for our purposes here, we avoid complicating our computations by incorporating this effect.

\begin{figure}[h!]
\begin{minipage}{.5\linewidth}
\centering
\subfloat{\includegraphics[scale=.75]{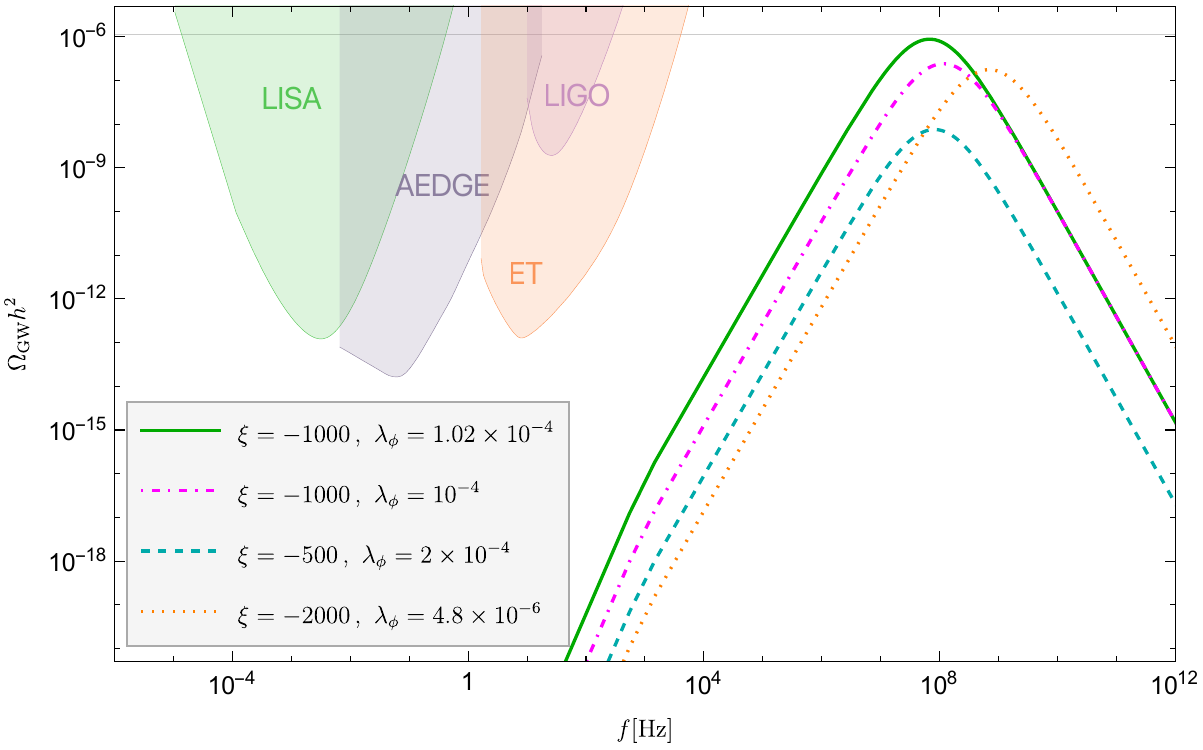}}
\end{minipage} \par\medskip
\begin{minipage}{.5\linewidth}
\centering
\subfloat{\includegraphics[scale=.75]{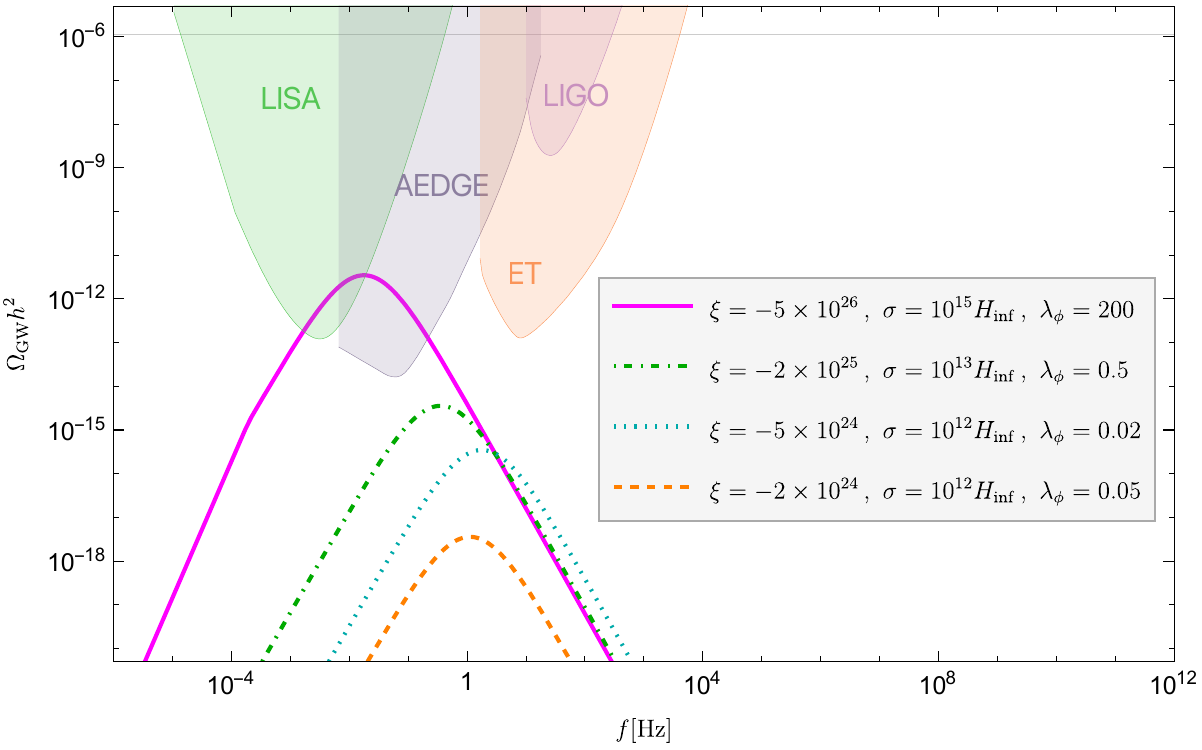}}
\end{minipage}
\caption{Gravitational wave spectra from phase transitions for $\xi_{\phi}<0$ according to Fig. \ref{fig:V-xi-negative}. The scale of reheating is set at $H_{\rm reh} \approx 9 \times 10^{-15}$ GeV with high scale inflation $H_{\rm inf}=10^{12}$ GeV and $\sigma = H_{\rm inf}$ (top) or low scale inflation $H_{\rm inf}=10^{-8}$ GeV (bottom). The BSM couplings are tuned accordingly in each case, so that they boost the GW signal while maintaining $C \approx-5.4$ (defined in Eq. (\ref{eq:delta_C})), and complying with the constraint (\ref{eq:inf_bound}).}
\label{fig:GW_xi_negative}
\end{figure}

\begin{figure}[h!]
\begin{minipage}{.5\linewidth}
\centering
\subfloat{\includegraphics[scale=.75]{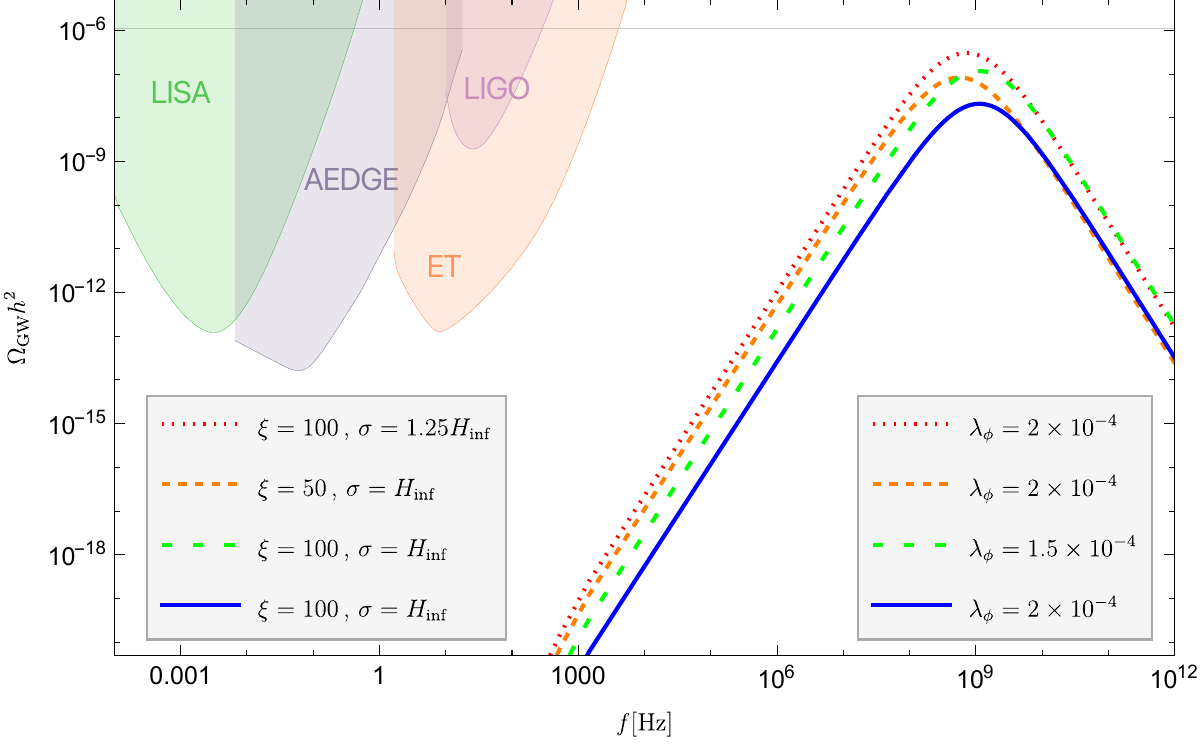}}
\end{minipage} \par\medskip
\begin{minipage}{.5\linewidth}
\centering
\subfloat{\includegraphics[scale=.75]{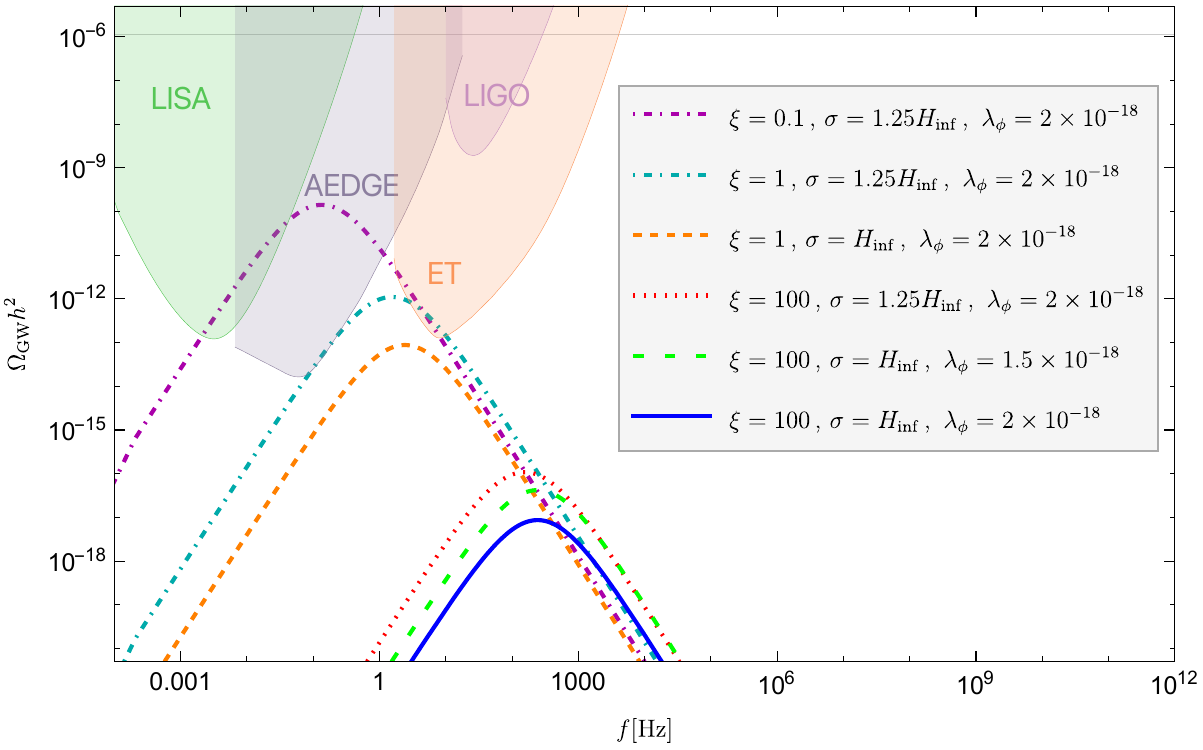}}
\end{minipage}
\caption{Gravitational wave spectra from phase transitions for $\xi_{\phi}>0$ according to Fig. \ref{fig:V-xi-positive}. The reheating scale is fixed at $H_{\rm reh} \approx 9 \times 10^{-15}$ GeV and the scale of inflation at $H_{\rm inf}=10^{12}$ GeV (top) and $H_{\rm inf}=10^{-8}$ GeV (bottom). The dark scalar's couplings have been chosen so that sufficiently strong FOPTs without supercooling are produced, with boosted GW signals that do not violate the BBN bound.}
\label{fig:GW_xi_positive}
\end{figure}

The resulting GW spectra from the phase transitions described in Sec. \ref{sec:2} for negative non-minimal coupling are shown in Fig. \ref{fig:GW_xi_negative} and for $\xi_{\phi}>0$ in Fig. \ref{fig:GW_xi_positive}, considering the most relevant region of the parameter space of the BSM couplings, according to the constraints presented in Sec. \ref{sec:2-3}. We have included two plots in each figure, with the upper one corresponding to signals originating from typical high inflationary scales $H_{\rm  inf}=10^{12}$ GeV, and the lower one to settings where the scale of inflation approaches the EW scale $H_{\rm  inf}=10^{-8}$ GeV. Since $H_{\rm col} \sim H_{\rm inf}$, this shows the range of GW signatures between the typically high frequency spectra associated with the high energy scales of the early universe, and lower inflationary scales that could be considered, as they are not a priori ruled out. However, note that the lowest scales that can be accommodated by realistic quintessential-like inflationary models are around $10^8$ GeV. Therefore, the bottom plots in Figs. \ref{fig:GW_xi_negative} and \ref{fig:GW_xi_positive} act as illustrative extreme examples that allow us to visualise approximately the interpolation of GW spectra from higher scales. Also, it should be pointed out that the energy density released by GWs cannot exceed the value of $\Omega_{\rm BBN} h^2 = 1.12 \times 10^{-6}$, because it would affect the Big Bang Nucleosyntheis (BBN). The BBN bound is represented in the plots by the thin horizontal grey line. We have included the detector sensitivities of LISA \cite{Caprini:2019pxz, Bartolo:2016ami}, AEDGE \cite{AEDGE:2019nxb}, the Einstein Telescope (ET) \cite{Punturo:2010zz, Hild:2010id} and LIGO \cite{LIGOScientific:2019vic, LIGOScientific:2016fpe, LIGOScientific:2014pky, Thrane:2013oya} to depict the observational prospects of such signatures with current and future experiments.

In each computation of a GW spectrum, the combination of the BSM couplings was chosen accordingly so that it provides the strongest possible GW signal, without crossing the BBN bound or involving supercooling, while also ensuring that bubble percolation is completed sufficiently quickly so that the background is approximately static. The low reheating scale, which allows for a curvature-induced EW symmetry breaking, enhances the GW spectra via the corresponding denominators in Eq. (\ref{eq:GW}) and limits decisively the viable parameter space of the dark scalar's couplings. It is worth highlighting that to compute the GW spectra from PTs where $\xi_{\phi}<0$, which can be reliable without performing detailed numerical bubble simulations, we limited ourselves to the area of the parameter space where $C \approx - 5.4$, meaning that $\lambda_{\phi}^{-1} \approx 10 |\xi_{\phi}| \left( \frac{H_{\rm inf}}{\sigma}\right)^2$ according to Eq. (\ref{eq:delta_C}). With this choice the potential maintains always an appropriate double-well shape, that allows the phase transition to proceed only via tunnelling. For lower values of $C$, the potential's symmetry would be restored, as shown in Fig. \ref{fig:GW_xi_negative} for $C=-10$. For higher values, the vacuum at $\phi_2$ would not evolve from a tv to a fv that could tunnel towards $\phi_1=0$, and thus a FOPT would not occur.

\clearpage

\section{Conclusions} \label{sec:4}

In this work, we examined whether gravitational waves can be generated from a curvature-induced first-order phase transition of a non-minimally coupled scalar field, that plays the role of dark matter, with a portal to the Higgs field. This was investigated in the context of a dynamical spacetime during the transition from inflation to kination, while also considering the possibility for inducing electroweak symmetry breaking in this manner. We explored a wide range of inflationary scales and both cases of positive and negative values for the non-minimal coupling, while taking into account the phenomenological constraints imposed on the BSM model's couplings from observations and experiments. The resulting GW amplitudes are boosted by extended periods of kination, akin to the quintessential-like inflationary models, and carry a characteristic tilt on their lower-frequency tail that serves as evidence of the alternative expansion history after inflation.

More specifically, we analysed the conditions for the realisation of a strong FOPT in vacuum, which led to the generation of a stochastic GW background from the collisions of the dark scalar's true-vacuum bubbles. Similarly as in Ref. \cite{Kierkla:2023uzo}, the phase transition complete approximately instantaneously with respect to the cosmological background, its strength depends on the total difference of the potential energy between the two vacua, but its inverse duration is sensitive to the evolution of the corresponding Euclidean action in each case of $\xi_{\phi}$. We applied the recent result of Ref. \cite{Matteini:2024xvg}, using the 1-loop result for the decay rate in vacuum, which can interpolate between regimes of thin and thick bubble walls and is thus valid for the PTs of both signs of the non-minimal coupling. The peak frequency and amplitude of the generated GWs depend on the ratio of the Hubble scales between bubble percolation and reheating. In particular, the extremely low reheating scale of $H_{\rm reh} \sim 10^{-14}$ GeV, corresponding to the assumption of instantaneous thermalisation with temperature $T_{\rm reh} \leq 80$ GeV, enhanced the GW signal so much that it strongly confined the parameter space of the BSM couplings that could comply with the BBN bound. For the specified choices of couplings depicted in Figs. \ref{fig:GW_xi_negative} and \ref{fig:GW_xi_positive}, it is possible to obtain GW spectra that could fall within the reach of future detectors. Although realistic high inflationary scales lead to peak frequencies of GW spectra that are higher than those probed by the upcoming experiments, the origin of such signatures is indisputably linked with the early universe, since there is convolution with other sources at this frequency range. In addition, we can readily distinguish the almost symmetric GW spectra obtained here from the standard paradigm of thermal phase transitions, where the interaction of the bubbles with the surrounding plasma (sound waves, turbulence, etc.) leads to enhanced high-frequency tails \cite{LISACosmologyWorkingGroup:2022jok, Cembranos:2024pvy}.

The BSM theory considered in this study expands the self-interacting model of Ref. \cite{Cosme:2018wfh}, by introducing a cubic term to the dark scalar potential. This allowed for the realisation of a strong PT due to the evolution of the curvature scalar, which was coupled to the scalar field, according to the mechanism developed in Ref. \cite{Kierkla:2023uzo}. Although we found agreement between the case of $\xi_{\phi}>0$ and the behaviour observed in Ref. \cite{Kierkla:2023uzo}, the additional case of $\xi_{\phi}<0$ had qualitatively different dynamics. This complicated the analysis especially when considering the subdominant role that the DM scalar was expected to have during inflation, which placed a firm constraint on the couplings according to Eq. (\ref{eq:inf_bound}). The presence of the Higgs-portal term was irrelevant for most of the cosmological evolution explored here, since the EW symmetry breaking, and therefore the Higgs' acquisition of its non-zero vev, was postponed to much later times with a very low reheating temperature $T_{\rm reh} \leq 80$ GeV. This approach followed the reasoning of Ref. \cite{Cosme:2018wfh}, so that the spontaneous breaking of the EW symmetry is not induced by the decreasing temperature of the primordial plasma. However, in contrast to Ref. \cite{Cosme:2018wfh}, EWSB was not induced by the rolling of the dark scalar down its potential towards the origin, but due to the decrease of spacetime curvature in the post-inflationary universe. This is viable only when the Higgs-portal coupling is extremely weak, as shown by Eq. (\ref{eq:g-bound}), since the large vev of the dark scalar could affect symmetry breaking. Coincidentally, the same coupling has to be extremely weak to allow the BSM scalar to act as DM after EWSB according to Eq.(\ref{eq:g_DM}),
\begin{align}
    10^{-32} \lesssim g \lesssim 10^{-21} \, , \qquad 10^{-30} \textrm{ GeV} \, \lesssim m_{\phi} \lesssim 10^{-19} \textrm{ GeV} \, ,
\end{align}
for the relevant parameter space across the inflationary scales considered
\begin{align}
    -10^3 \lesssim \xi_{\phi} \lesssim 10^2 \, , \qquad \sigma \sim H_{\rm inf} \, , \qquad \lambda_{\phi} \lesssim 10^{-4} \, ,
\end{align}
which trivially satisfies the bounds (\ref{eq:g_branching_ratio_bound}) and (\ref{eq:g_radiative}). This implies that when a non-minimally coupled Higgs-portal scalar field with a renormalizable potential plays the role of very light dark matter in the late universe, it allows for the breaking of the EW to be induced by the evolution of spacetime curvature, provided that reheating takes place at sufficiently low temperatures.

In any case, we believe that it is quite interesting that it is possible to identify a new observational signature of a Higgs-portal dark scalar field through its non-minimal coupling to curvature. Even though the prospects for the detection of the associated gravitational waves are somewhat unfavourable, it is exciting that they might be, at least theoretically, putative messengers of the properties of an interesting class of theories involving inflation and a scalar field DM candidate that can itself undergo a first order phase transition and have an effect on the spontaneous breaking of the EW symmetry. This naturally leads to a promising and multi-faceted outlook for future investigations. The framework developed here could be applied to different BSM extensions with various field content and couplings. Within the same context, different inflationary models can be explored, where understanding the transition from inflation/kination to the radiation-dominated epoch through reheating being particularly crucial. At the same time, we expect that it would be fruitful to extend these calculations to theories including non-minimal couplings of matter to curvature \cite{Bertolami:2017svl, Gomes:2016cwj, Bertolami:2010ke}, given the pronounced role that spacetime curvature can play in scenarios like the ones explored here. These endeavours could offer valuable tools and insights that can enhance our understanding of the early universe, complimentary to accelerator experiments and observational surveys.

\section*{Acknowledgements}
We would like to thank Marek Lewicki, Mar Bastero-Gil, António P. Morais, and Marco Finetti for the useful discussions and suggestions, and Maciej Kierkla for his contribution in the graphic representation of the GW spectra. This work was supported by FCT - Fundação para a Ciência e Tecnologia, I.P. through the project CERN/FIS-PAR/0027/2021, with DOI identifier 10.54499/CERN/FIS-PAR/0027/2021.

\appendix
\section{Self-interacting scalar field with a Higgs portal} \label{app:1}

\begin{figure}[h!]
    \centering
    \includegraphics[scale=0.9]{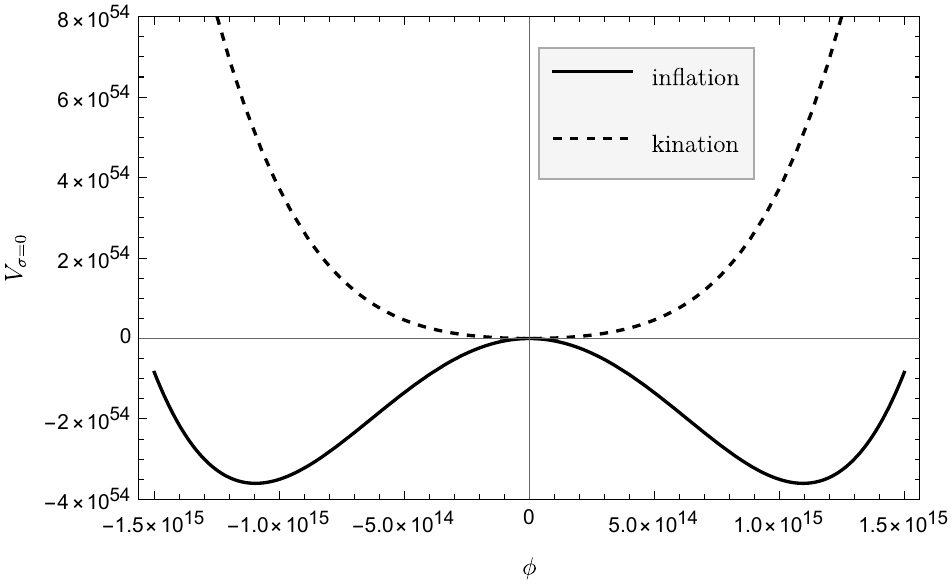}
    \caption{Evolution of the dark scalar potential from inflation (solid) to kination (dashed), given by Eq. (\ref{eq:V-potential}) for $\xi_{\phi}=-1 \, , \, \, \sigma=0$ and $\lambda_{\phi}=10^{-5}$.}
    \label{fig:V_sig0}
\end{figure}

If we set $\sigma=0$ in the dark scalar potential (\ref{eq:V-potential}), we recover the self-interacting model from Ref. \cite{Cosme:2018wfh}, with its evolution from inflation to kination shown in Fig. \ref{fig:V_sig0}. For $\xi_{\phi}<0$, there are two degenerate vacua at $\phi = \pm \sqrt{\frac{|\xi_{\phi}| R}{\lambda_{\phi}}-\frac{g^2 h^2}{2 \lambda_{\phi}}}$ that become shallower and move closer to the origin during the transition from inflation to kination, eventually resulting in a single global minimum at $\phi=0$. This behaviour is reversed for $\xi_{\phi}>0$ albeit not explored in Ref. \cite{Cosme:2018wfh}. During inflation, the large vev of the spectator scalar field, $\phi_{\rm inf} = \sqrt{\frac{12 |\xi_{\phi}| H^2_{\rm inf}}{\lambda_{\phi}}}$, generates a large contribution to the effective mass of the Higgs field through the portal, $m_h^{\rm eff} \propto  \frac{g}{\sqrt{2}}\phi_{\rm inf} = \sqrt{\frac{6 |\xi_{\phi}|}{\lambda_{\phi}}} g H_{\rm inf}$. This keeps the Higgs field at the origin, by suppressing its de Sitter fluctuations $ \langle h^2 \rangle \propto H_{\rm inf} / m_h^{\rm eff}$ for the appropriate couplings $g / \sqrt{\lambda_{\phi}} \sim 10^2$ and $\xi_{\phi} < - 0.1$ \cite{Cosme:2018wfh}. After symmetry restoration of the potential during kination, the scalar rolls down towards the minimum and induces EW symmetry breaking when crossing the critical value $\phi_{\rm crit} = \sqrt{2 \lambda_h} (v / g)$ according to Eq. (\ref{eq:h_EW}). Consequently, the Higgs field relaxes around its vev as the oscillations of the BSM scalar around $\phi=0$ slow down and spacetime curvature tends to approximate flatness. However, note that this is true only if the reheating temperature is $T_{\rm reh} \leq 80$ GeV, to avoid thermal restoration of the EW symmetry.

\section{Potential redefinition for a phase transition from a positive true vacuum} \label{app:2}

The evolution of the BSM potential for $\xi_{\phi}<0$ in Fig. \ref{fig:V-xi-negative} implies that the phase transition takes place from a positive true vacuum to the false vacuum located at the origin. However, the prescription for calculating the decay rate and the PT parameters in Sec. \ref{sec:3} is formulated for trajectories starting from a fv that is firmly fixed at the origin towards either a positive \cite{Adams:1993zs} or a negative \cite{Matteini:2024xvg} tv. Therefore, in order to accurately calculate the dynamics of the PT and the corresponding GW spectra, we have to transform the potential (\ref{eq:V-potential}) accordingly as
\begin{multline}
        W (\phi, h=0, R) = V(\phi+\phi_2, h, R) - V(\phi_2, h, R) = \\
        = \left[ \frac{\sigma^2}{4 \lambda_{\phi}} \left(1 + \sqrt{1 + \frac{4 \lambda_{\phi} |\xi_{\phi}| R}{\sigma^2}}\right) + |\xi_{\phi}| R \right] \phi^2 - \frac{|\sigma|}{2} \left( \frac{1}{3} + \sqrt{1 + \frac{4 \lambda_{\phi} |\xi_{\phi}| R}{\sigma^2}} \right) \phi^3 + \frac{\lambda_{\phi}}{4} \phi^4 \, .
        \label{eq:W_potential}
\end{multline}
Note that in order to match with the framework of this study, where the tv lies at positive values, the cubic coupling must have the opposite sign $\sigma<0$. However, as long as we are consistent, this does not affect the dynamics and GWs of the PT, it is only a matter of formalism and convenience for the calculation of the key quantities of the PT. Following the same steps as in Sec. \ref{sec:2}, the potential (\ref{eq:W_potential}) is written in the reduced dimensionless form as
\begin{align}
    \Tilde{W}(\chi, \delta) = \frac{\Tilde{\delta}}{2}\chi^2 - \chi^3 + \frac{1}{4} \chi^4 \, ,
\end{align}
where the field redefinition and its corresponding quadratic coupling are given by
\begin{align}
\chi (R) = \frac{2 \lambda_{\phi}}{|\sigma|} \left( \frac{1}{3} + \sqrt{1 + \frac{4 \lambda_{\phi} |\xi_{\phi}| R}{\sigma^2}} \right)^{-1} \phi \, ,
\end{align}
\begin{align}
\Tilde{\delta}(h=0, R) = \left( \frac{8 \lambda_{\phi}}{\sigma^2 } \right) \frac{ \frac{\sigma^2}{4 \lambda_{\phi}} \left(1 + \sqrt{1 + \frac{4 \lambda_{\phi} |\xi_{\phi}| R}{\sigma^2}}\right) + |\xi_{\phi}| R }{\left( \frac{1}{3} + \sqrt{1 + \frac{4 \lambda_{\phi} |\xi_{\phi}| R}{\sigma^2}}\right)^2} = \frac{2 \left( 1 - \frac{4 \delta}{9} + \sqrt{1 - \frac{4 \delta}{9}}\right)}{\left( \frac{1}{3} + \sqrt{1 - \frac{4 \delta}{9}}\right)^2}\, ,
\label{eq:deltaW}
\end{align}
where $\delta<0$ during inflation according to Eq. (\ref{eq:delta}), but $\Tilde{\delta}>0$. The overall rescaling of the potential is $ W = \left[\frac{|\sigma|}{2} \left( \frac{1}{3} + \sqrt{1 + \frac{4 \lambda_{\phi} |\xi_{\phi}| R}{\sigma^2}} \right) \right]^4 \lambda_{\phi}^{-3} \Tilde{W}$, and its corresponding extrema lie at
\begin{align}
   \phi_1^W &= 0 \,, \qquad  \qquad \qquad \quad \qquad \qquad \qquad \quad \,  \chi_1 = 0 \, , \label{eq:chi1} \\
    \phi_2^W (h, R) &= \frac{|\sigma|}{2\lambda_{\phi}}\left(1+\sqrt{1 + \frac{4 \lambda_{\phi} |\xi_{\phi}| R }{\sigma^2}}\right) \,, \quad \chi_2 (\Tilde{\delta}) = \frac{3}{2} \left( 1 + \sqrt{1 -\frac{4 \Tilde{\delta}}{9}} \right) \, , \label{eq:chi2} \\
   \phi_3^W (h, R) &= \frac{|\sigma|}{\lambda_{\phi}}\sqrt{1 + \frac{4 \lambda_{\phi} |\xi_{\phi}| R}{\sigma^2}} \,, \quad \qquad \qquad \chi_3 (\Tilde{\delta}) = \frac{3}{2} \left( 1 - \sqrt{1 -\frac{4 \Tilde{\delta}}{9}} \right) \, , \label{eq:chi3} 
\end{align}
as long as $ 4 \delta / 9 > - 1$ or equivalently $ 4 \Tilde{\delta} / 9 < - 1$ . Finally, we highlight that the potential difference between the fv and tv of the potential,
\begin{align}
    \Delta W &= W\left(\phi_2^W\right) - W\left(\phi_1^W\right) = \frac{\sigma ^4}{96 \lambda ^3} \left(1 - \frac{6 \lambda  \xi_{\phi}  R}{\sigma^2} + \sqrt{1-\frac{4 \lambda  \xi_{\phi}  R}{\sigma ^2}}\right) \left( 1 + \sqrt{1-\frac{4 \lambda  \xi_{\phi}  R}{\sigma ^2}}\right)^2 = \nonumber \\
    &= \frac{\sigma ^4}{96 \lambda ^3} \left(1 - \frac{2 \delta}{3} + \sqrt{1-\frac{4 \delta}{9}}\right) \left( 1 + \sqrt{1-\frac{4 \delta}{9}}\right)^2 \, ,
\end{align}
is equivalent to (\ref{eq:DeltaV}). This is expected since the transformation we performed does not affect the relative depths of the potential vacua, but it is just a convenient redefinition of the potential that encapsulates the FOPT of Fig. \ref{fig:V-xi-negative}. Therefore, it is trivial that Eq. (\ref{eq:alpha-BSM}) holds in this case, since the transition strength is the same with either prescription. 

\bibliography{references}

\end{document}